\documentclass[floats,twocolumn,aps]{revtex4}
\usepackage{latexsym}
\usepackage{graphicx}
\newcommand\beq{\begin{equation}}
\newcommand\eeq{\end{equation}}
\newcommand\bea{\begin{eqnarray}}
\newcommand\eea{\end{eqnarray}}
\newcommand\non{\nonumber}
\newcommand\bib{\bibitem}

\begin{document}
\title{\bf Quantum simulation of Dirac fermion mode, Majorana fermion mode and Majorana-Weyl fermion
mode  
in cavity QED lattice}

\author{\bf Sujit Sarkar}
\address{\it Poornaprajna Institute of Scientific Research,
4 Sadashivanagar, Bangalore 5600 80, India.\\
e-mail: sujit.tifr@gmail.com \\
}
\date{\today}

\begin{abstract}
Quantum simulation aims to simulate a quantum system using a
controble laboratory system that underline the same mathematical model.
Cavity QED lattice system is that prescribe system to simulate the relativistic
quantum effect.
We quantum simulate the  
Dirac fermion mode, Majorana fermion mode and Majorana-Weyl fermion mode 
and a crossover between them in
cavity QED lattice.  
We also present the different 
analytical relations between the field operators for different mode excitations.
\end{abstract}
\maketitle

{\bf Introduction:} 
The difficulties in observing the real quantum relativistic effects 
have generated immense interest in the quantum simulation physics.
In recent years, there has been increased interest in the
simulation of relativistic quantum effects using different physical
system in which parameters tunibility allows access to different
physical regimes 
\cite{gerri,seo,kane,bardyn,bkumar,trif,sola1,sola2,xzhang,otterbach,clark}. 
These difficulties of observing quantum relativistic effects stimulate us to
study the quantum simulation physics
of Dirac fermion mode, Majorana fermion mode and Majorana-Weyl fermion modes 
in one dimensional
cavity QED lattice and a crossover from one mode of excitations
to the another using the tunibility of the physical parameters of
the system.\\
In quantum simulation, one aim is to simulate a quantum system using 
a controllable laboratory system that underlines the same mathematical
models. Therefore it is possible to simulate a quantum system that can
be neither efficiently simulated on a classical computer nor easily
accessed experimentally.\\
The recent experimental success in engineering strong interaction 
between the photons and atoms in high quality micro-cavities opens up
the possibility to use the light matter system as quantum simulators for
many body physics [9-29]. 
Many interesting results are coming out to understand the complicated 
quantum many body system. 
A focus
on the coupled cavities is one of the most potential candidate for
an efficient quantum simulator due to the control of the microcavities
parameters and success of fabrication of large scale cavity arrays [25-26].\\
The concept and existence of Majorana fermion mode is one of the most advance research area
in quantum condensed matter system.
Majorana introduced a special kind of fermions which are their
own antiparticle, i.e., the neutral particle \cite{majo,wil}. He introduced
this particle to describe neutrions. In recent years, there are
several candidates of Majorana fermions in quantum condensed
matter system like quantum Hall system with filling fraction 
$ 5/2 $ \cite{read,read2}. Kitaev first found the existence of Majorana 
fermion mode in one dimensional model \cite{kitaev}. Many research 
groups have already proposed the physically existence of
MFs at the edge state of 1D system like electrostatic defects lines
in superconductor, quasi-one dimensional superconductor and
cold atom trapped in one dimension \cite{wimmer,kop}.
Majorana fermions obey the non-Abelian statics both in 2D and
1D, allowing of certain gate operation required in quantum
computation \cite{nayak}.\\ 
In the present study one of our goal is to predict the presence
of Dirac fermion mode, Majorana fermion mode and 
Majoran-Weyl fermion mode and crossover in our model system.
In the Majorana-Weyl fermion mode, the fermion mode satisfy
the condition for particle and antiparticle condition but the
excitation is massless.\\ 
There are few studies to simulate the Majorana fermion modes in
cavity QED system but the model Hamiltonian and the associate
relevant physics are different and at the same time there is
no coossover study between the Dirac and Majorana fermion modes
in the literature of cavity QED system [9-11].\\  
The authors of Ref.3 have found the process of Dirac to Majorana fermion
converator on the surface of a 3D topological insulator. A Dirac fermion
injected by the voltage source is split into a pair of Majorana fermion 
and then finally recombine before going to drain. 
But in our present study there is no such split and fusion process of
Dirac and Majorana fermion. In our study the system is in Dirac
fermion mode when only a atom-photon coupling or a single Rabi frequency
oscillation present in the system and the existence of Majorana
fermions when two atom photon couplings and the two laser frequencies
are simultaneously present in the system. For a quantum
simulated Hamiltonian for Majorana fermion mode for a specific
mathematical relation between the Rabi frequencies oscillation, the
atom-photon coupling strengths and the laser field detuning.\\
To the best of our knowledge the quantum simulation physics for
the different kind of fermionic mode in a same cavity QED system
is absent in the literature.\\
In the present study, we simulate two model Hamiltonians
through the proper tuning of Rabi frequencies and the atom-photon
coupling strengths in the system which quantum simulate different 
modes of fermionic excitations.\\  
{\bf The Model Hamiltonian:\\}
The Hamiltonian of our present study consists of three parts:
\beq
H ~= ~ {H_A} ~+~ {H_C}~+~{H_{AC}}
\eeq     
The Hamiltonians are the following\\
\beq
 {H_A} ~=~ \sum_{j=1}^{N} { {\omega}_e } |e_j > <e_j | ~+~ 
{\omega}_{ab} |b_j > <b_j |  
\eeq
where $j$ is the cavity index. ${\omega}_{ab} $ and ${\omega}_{e} $ are 
the energies of the state $ | b> $ and the excited state respectively. The
energy level of state $ |a > $ is set as zero. $|a>$ and $|b> $ are
the two stable state of a atom in the cavity and $|e> $ is the
excited state of that atom in the same cavity. 
The following Hamiltonian describes the photons in the cavity, \\
 \beq
  {H_C} ~=~ {{\omega}_C} \sum_{j=1}^{N} {{a_j}}^{\dagger} {a_j} ~+~
{J_C} \sum_{j=1}^{N} ({{a_j}}^{\dagger} {a_{j+1}} + h.c ),  
\eeq 
where ${a_j}^{\dagger}({a_j})$ is the photon
creation (annihilation) operator for the photon field in the $j $'th cavity, ${\omega}_C $
is the energy of photons and $ J_C $ is the tunneling rate of photons
between neighboring cavities.
The interaction between the atoms and photons and also by the driving lasers
are described by \\
\beq
 {H_{AC}}~=~ \sum_{j=1}^{N} [ (\frac{{\Omega}_a}{2} e^{-i {{\omega}_a} t} +
{g_a} {a_j}) |e_j > < a_j | + h.c] + [a \leftrightarrow b ] .  
\eeq
Here ${g_a} $ and ${g_b} $ are the couplings of the cavity mode for the
transition from the energy states $ |a > $ and $ | b> $ to the excited state.
${\Omega}_a $ and ${\Omega}_b $ are the Rabi frequencies of the lasers
with frequencies ${\omega}_a $ and $ {\omega}_b $ respectively.\\
The authors of Ref. \cite{hart1,hart2,sujop}
have derived an effective spin model by considering the following physical
processes:
A virtual process regarding the emission and absorption of
photons between the two stable  states of neighboring cavity yields the resulting 
effective Hamiltonian as
\beq
{H_{xy}} = \sum_{j=1}^{N}  B {{\sigma}_j}^{z} ~+~\sum_{j=1}^{N} 
(\frac{J_1}{2} {{\sigma}_j}^{\dagger} {{\sigma}_{j+1}}^{-} ~+~
\frac{J_2}{2} {{\sigma}_j}^{-} {{\sigma}_{j+1}}^{-} + h.c )
\eeq 
When $J_2 $ is real then this Hamiltonian reduces to the XY model.
Where ${{\sigma}_j}^{z} = |b_j > <b_j | ~-~ |a_j > <a_j | $,
${{\sigma}_j}^{+} = |b_j > <a_j | $, ${{\sigma}_j}^{-} = |a_j > <b_j | $ .
\beq
H_{xy} 
 =  \sum_{i=1}^{N} B ( {{\sigma}_i}^{z}~+~{J_x} {{\sigma}_i}^{x}
{{\sigma}_{i+1}}^{x} ~+~ {J_y} {{\sigma}_i}^{y}
{{\sigma}_{i+1}}^{y}) .
\eeq
With ${J_x} = (J_1 + J_2 ) $ and ${J_y} = (J_1 - J_2 ) $.\\
We follow the references \cite{james,hart1}, to present the analytical 
expression for the different physical parameters of the system.\\
$ B = \frac{\delta_1}{2} - \beta $, 
$\beta$ is define in Ref. \cite{beta}.
\beq
{J_1} = \frac{\gamma_2}{4} ( \frac{{|{\Omega_a}|}^2 {g_b}^2 }{{ {\Delta}_a }^2 }
 +  \frac{{|{\Omega_b}|}^2 {g_a}^2 }{{ {\Delta}_b }^2 } ) ,
{J_2} = \frac{\gamma_2}{2} ( \frac{{\Omega_a} {\Omega_b} g_a g_b }{{\Delta}_a {\Delta_b} }
 ).
\eeq
Where 
$ \gamma_{a,b} = \frac{1}{N} \sum_{k} \frac{1}{ {\omega}_{a,b} - {\omega}_k } $
$ \gamma_{1} = \frac{1}{N} \sum_{k} \frac{1}{ ( {\omega}_{a}+  {\omega}_{b})/2 - {\omega}_k } $ and
$ \gamma_{2} = \frac{1}{N} \sum_{k} \frac{e^{ik} }{ ( {\omega}_{a}+  {\omega}_{b})/2 - {\omega}_k } $
${\delta_1} = {\omega}_{ab} - ({\omega}_a - {\omega}_b )/2 $, 
${\Delta}_a = {\omega}_e - {\omega}_a$.  
${\Delta}_b = {\omega}_e - {\omega}_a -({\omega}_{ab} - {\delta_1})$.
${{\delta}_a}^{k} = {\omega}_e - {\omega}_k $,
${{\delta}_b}^{k} = {\omega}_e - {\omega}_k  -({\omega}_{ab} - {\delta_1}) $,
${\omega}_k = {\omega}_c + J_c \sum_{k} cosk $.
$g_a$ and $g_b$ are the couplings of respective transition to the cavity mode,
${\Omega}_a $ and ${\Omega}_b$ are the Rabi frequency of laser with frequency
$\omega_a $ and $\omega_b $. \\

{\bf Quantum Simulation for Dirac Fermion mode and Mathematical Relation Between the Fields :}\\
Here we quantum simulate the Dirac fermion physics through the proper
tuning of cavity QED lattice parameters.
This condition can be achieve when $J_x = J_y$. This condition implies
that $ 2 J_2 =0 $. It is clear from the analytical expression of $J_1 $
and $J_2 $ that to satisfy the condition one of the atom-photon 
coupling strength, i.e., $ g_a$ or $g_b$ be zero or one of the Rabi 
frequency oscillation ${\Omega}_a $ or ${\Omega}_b $ should be zero.
The analytical expression for $J_1 $ become 
$ J_1 = \frac{\gamma_2}{4} \frac{ {|{\Omega_a }|}^2 {g_b}^2 }{ {\Delta_a}^2} $ or
$ J_1 = \frac{\gamma_2}{4} \frac{ {|{\Omega_b }|}^2 {g_a}^2 }{ {\Delta_b}^2} $.
The other condition is $ {\delta_1 } = 2 \beta $, i.e.,
$ 2 {\omega}_{ab} - ({\omega}_a - {\omega}_b )  = 4 \beta $.
Therefore the condition for quantum simulation relates upto the microscopic level.\\ 
In this limit the Hamiltonian reduced to
\begin{equation}
H = J \sum_{i =1 }^{N} ( {\sigma_i}^{x} {\sigma_{i+1}}^{x} + {\sigma_i}^{y} {\sigma_{i+1}}^{y} ) 
\end{equation}
After the Jordan-Wigner transformation and Abelian Bosonization study
one can write the above Hamiltonian as \cite{gia,frad}\\
\beq
H_0 = \sum_{s} \int \frac{dk}{2 \pi} \epsilon (k) {\psi_s}^{\dagger} (k) 
{\psi_s} (k) 
\eeq
\beq
H_0 = \sum_{s} \int_{-\Lambda}^{\Lambda} \frac{dk}{2 \pi} (k v_F )
( {{\psi}_{s,R}}^{\dagger} (k) {{\psi}_{s,R}} (k) -
{{\psi}_{s,L}}^{\dagger} (k) {{\psi}_{s,L}} (k) ). 
\eeq
Here we use $ {\epsilon}_k = k v_F $ near the Fermi points and
$ { {\psi_{s} }}^{\dagger} (x) = ( { {\psi}_{s,R}}^{\dagger} (x), 
{ {\psi}_{s,L}}^{\dagger} (x) ) $.
Following the discussions in addendum, we can write the above Hamiltonian
in the following form of Dirac equation without any mass term.\\
\beq
H = 2 J \int dx \bar{\psi} (i {\gamma_1} ) {\partial_x} {\psi}
\eeq
and $ {\psi}^{\dagger} (x) = ( {{\psi}_R}^{\dagger}, {{\psi}_L}^{\dagger} ) $.
$ { \bar{\psi}}^{\dagger} (x) = ( {{\psi}_L}^{\dagger}, {{\psi}_R}^{\dagger} ) $,\\
$ \psi_s (x) = \int \frac{dk}{2 \pi} {\psi_s }(k) e^{ikx} $.
It is customery to introduce the Dirac matrices in Dirac equation. The ${\gamma}$
matrices are the following.
$ {\gamma}^{0} = \left (\begin{array}{cc}
      0 & 1 \\
    1  & 0
        \end{array} \right ) $ ,
$ {\gamma}^{1} = -i {\sigma}_y =  \left (\begin{array}{cc}
      0 & -1 \\
    1  & 0
        \end{array} \right ) $ ,
$ {\gamma}^{5} = {\gamma_0} {\gamma_1} = \sigma_z =  \left (\begin{array}{cc}
      1 & 0 \\
    0  &  -1 
        \end{array} \right ) $ , \\
${\psi}_R $ and ${\psi}_L $ are the fermionic field for the right and
left movers electron.
The scalar field and its dual can be expressed as
$ \phi (x) = \phi_R (x) + \phi_L (x) $, $\theta (x) = - {\phi}_R (x) + {\phi}_L (x) $.
The detail derivation of the analytical relation between the wave function and
the scalar field are relegated to the addendum. Here we only present the final form
Now we present the mathematical relation between the operators, $ e^{\pm \phi (x)} $ and 
$ e^{\pm \theta (x)} $. \\
\begin{equation}
e^{i \alpha \sqrt{\pi} \phi (x1 )} e^{i \beta \sqrt{\pi} \theta (x2 )} =  
e^{i \beta \sqrt{\pi} \theta (x2 )} e^{i \alpha \sqrt{\pi} \phi (x1 )} ~sign (x2 -x1)   
\end{equation}
\begin{equation}
e^{i \alpha \sqrt{\pi} \phi (x1) } {\psi_{R/L}} (x2) = {\psi_{R/L}} (x2) 
e^{i \alpha \alpha \sqrt{\pi} \phi (x1) }~sign(x2 - x1) 
\end{equation} 
\begin{equation}
e^{i \beta \sqrt{\pi} \theta (x1) } {\psi_{R/L}} (x2) = - {\psi_{R/L}} (x2) 
e^{i \beta \sqrt{\pi} \theta (x1) }~sign(x2 - x1) . 
\end{equation} 
Therefore it is clear from our theoretical analysis that the cavity QED lattice
shows the Dirac fermion like mode when one of the atom-photon coupling ( $g_a $ or
$g_b$ ) is zero or one of the applied Rabi frequencies ( ${\Omega}_a $, ${\Omega}_b $)
is zero. \\ 
 
{\bf Dirac Equation for Majorana Fermion mode and Condition of Majorana-Weyl fermionic mode:}\\
Here we quantum simulate the Majorana fermion mode and the Majorana-Weyl fermion mode.
Majorana thought whether it might be possible for spin-1/2 particle
to be its own anti-particle. To get an equation alike to Dirac
equation but capable of governing a real field that requires the
$\gamma $ matrices of that equation must satisfy the Clifford algebra
are purely imaginary. Here we show explicitly that the
$\gamma $ matrices of Majorana equation are purely imaginary and field
satisfy the particle-antiparticle equivalent condition.\\ 
We also derive the condition for the excitation of Majorana-Weyl fermion
mode where the simulated mode shows the gapless excitation.\\ 
We consider $ J_1 = J_2 $,  
$J_x $ become $ J_1 + J_2 $ and  $ J_y =0 $. 
In the microcavity array, the condition for $J_1 = J_2 $ achieve when
\beq
{{\Omega}_a}^2  {g_b}^2 {\Delta_b}^2 + {{\Omega}_b}^2  {g_a}^2 {\Delta_a}^2 =
2 {\Omega}_a {\Omega}_b  g_a g_b {\Delta}_a {\Delta}_b . 
\eeq
The above condition implies that $ {\Omega}_a = {\Omega}_b  \frac{g_a \Delta_a}{g_b \Delta_b }$.
The only constraint is that ${\Delta}_a \neq {\Delta}_b $, the magnetic field 
diverge when ${\Delta}_a = {\Delta}_b $. At the same time, ${\Omega}_a = {\Omega}_b $ and
$g_a = g_b$ are also not possible because this limit also leads to the condition
${\Delta}_a = {\Delta_b} $. Suppose we consider,
${\Omega}_a  = \alpha_1 {\Omega}_b $, $g_a = \alpha_2 g_b $ and ${\Delta}_a = \alpha_3 {\Delta}_b $.
These relations implies that 
$ {\alpha_1 }^2 + {\alpha_2 }^2 {\alpha_3 }^2 = 2 {\alpha}_1 {\alpha}_2 {\alpha}_3 $.
${\alpha}_1 = {\alpha}_2  {\alpha}_3  $, 
${\alpha}_1 , {\alpha_2 }$ and ${\alpha_3}$ are the
numbers. These analytical relations help to implement 
the transverse Ising model Hamiltonian but $\alpha_1 $, $\alpha_2 $ and $\alpha_3 $ should
not be equal to 1.\\
The quantum state engineering of cavity QED is in the state of art due to the
rapid progress of technological development of this field [1]. Therefore one can 
achieve this limit to get the desire quantum state. One can write the final
Hamiltonian as, 
\beq
H_{T} = B \sum_{j=1}^N ( {{\sigma}_z} (j) + \lambda {{\sigma}_x} (j) {{\sigma}_{x}} (j+1) ),
\eeq
where $\lambda = \frac{ J_1 + J_2}{B} $. 
The effective Hamiltonian become
the transverse Ising model which studied in the previous literature \cite{ss,mussardo,druff}. 

One can also write the starting Hamiltonian as through the rotation in spin basis\\
\beq
H = \sum_{n} [\lambda \sigma_z (n) \sigma_z (n+1) + s_x (n)].
\eeq
We recast the Hamiltonian in the following form because it will help us to
use the order and disorder operator directly to the derivation of the equation
of motion and finally the Dirac equation for Majorana 
fermion.\\ 
Here our main
motivation is to use some of important results of this model Hamiltonian 
to discuss the relevant physics of array of cavity QED system.
We introduce the order and disorder operator in the addendum 
\cite{mussardo,druff}. 
These operators are defining the sites of the dual lattice, i.e., we
define the operator between the nearest-neighbor site of the original
lattice.  
Here we define the Dirac spinor, 
$ \chi_1 (n) =  \sigma_z (n) \mu_z (n+ 1/2)$ and 
$ \chi_2 (n) =  \sigma_z (n) \mu_z (n- 1/2)$.\\ 
Now our main task is to find the equation of motion for the operators,
${\sigma}_3  (n) $ and ${\mu}_3 (n) $ which help us to build the 
Dirac equation. The detail derivation
are relagated to the appendix.\\
The equation of motion for the ${\sigma}_z (n)$ is the following:\\
\beq
\frac{\partial \sigma_z (n)}{\partial \tau} = [H, \sigma_z (n)]= {\sigma_x (n) \sigma_z (n) }
\eeq 
The equation of motion for $\mu_z (n+1/2) $ is the following:\\
\bea
\frac{\partial \mu_z (n+1/2)}{\partial \tau} & = & 
\lambda {\mu_x (n +1/2 ) \mu_z (n + 1/2) } \non\\
& & = \lambda \sigma_z (n) \sigma_z (n + 1/2) \mu_z (n + 1/2) \\  
\eea
\beq
\frac{ \partial \chi_1 (n) }{d \tau} = - \chi_2 (n) 
+ \lambda \chi_2 (n + 1). 
\eeq
\beq
\frac{ \partial \chi_2 (n) }{d \tau} = -\chi_1 (n) 
 + \lambda \chi_1 (n-1). 
\eeq
These two fields, $\chi_1 (n)$ and $\chi_2 (n) $ satisfy the following relations,
$ \{ \chi_1 (n1) , \chi_2 (n2) \} = 2 \delta_{n1, n2} $. One can write down
the above equation in the following compact form, \\
\beq
( {\gamma}^0  \frac{\partial }{\partial t} + {\gamma}^{3} \frac{\partial}{\partial r}
+ m ) \chi =0 . 
\eeq
where ${\chi}^{\dagger} = ( \chi_1 , \chi_2 )$
and $ m= \frac{1 - \lambda}{\alpha} $, 
$ {\gamma}^{0} = \left (\begin{array}{cc}
      0 & 1 \\
    1  & 0
        \end{array} \right ) ,
 {\gamma}^{3} = \left (\begin{array}{cc}
      1 & 0 \\
    0  & -1
        \end{array} \right ) $
.\\
One can also write the above Majorana equation in a compact form:\\
\beq
( i {\tilde{\gamma}}^{\mu} {\partial}_{\mu}  -m ) {\chi} =0. 
\eeq
Where 
${\tilde {\gamma}}^{0} = \left (\begin{array}{cc}
      0 & i \\
    i  & 0
        \end{array} \right ) ,
{\tilde {\gamma}}^{3} = \left (\begin{array}{cc}
      i & 0 \\
    0  & -i
        \end{array} \right ) $
.\\
Therefore we prove that the spinor field satisfy the Majorana condition of spin-1/2
particle and also the $\tilde{ \gamma} $ matrices are imaginary.\\
The condition for the massless Majorana fermion field is that $m =0, i.e., \lambda =1 $.
In this quantum simulation process, one can get through this massless excitation through
this analytical relation.\\
$ \frac{\gamma_2}{2} ( \frac{{|{\Omega_a}|}^2 {g_b}^2 }{{ {\Delta}_a }^2 }
 +  \frac{{|{\Omega_b}|}^2 {g_a}^2 }{{ {\Delta}_b }^2 } ) 
= \frac{\omega_{ab}}{2} - \frac{1}{4} (\omega_a - \omega_b ) -\beta $.\\
We term this fermionic mode as Majorana-Weyl fermionic mode.
In this study we obtain the different kind of fermionic modes
from the quantum simulation of the system, here 
there is no splitting of Dirac fermion mode into the 
Majorana fermion mode or Majorana-Weyl fermionic mode. 
There are no studies in the previous literature of
cavity QED where one has found a crossover from Dirac fermion modes to Majorana
fermion modes [9-11].\\ 
 
Here we present the analytical relations between the 
Majorana fermion operators with the order
and disorder operator with the free Dirac field in Abelian bosonization
theory.
Then one obtains the following sets of commutation
relations.\\
\beq
{\sigma}_z ( x1 ) {\mu}_z (x2) = {\mu}_z (x2) {\sigma}_z (x1 ) sign( x1 -x2)  
\eeq
\beq
{\sigma}_z ( x1 ) {\chi} (x2) = {\chi} (x2) {\sigma}_z (x1 ) sign( x1 -x2)  
\eeq
\beq
{\mu}_z ( x1 ) {\chi} (x2) = - {\chi} (x2) {\sigma}_z (x1 ) sign( x1 -x2)  
\eeq
It is very clear from the above analytical relations that
${{\chi}_1}^{\dagger}= {\chi}_1 $ and ${{\chi}_2}^{\dagger} = {{\chi}_2 }$.
The detail derivation is relegated in the appendix.
The above relation has similarity with the free Dirac field in Abalian
bosonization theory, where Dirac field operator is a local
product of two phase exponential depending on the scalar
field and its dual \cite{gia,gogo,frad}, as one study the
Luttinger liquid physics in Abelian bosonization theory.\\ 
{\bf Conclusions}\\
We have presented the existence of Dirac fermion mode, 
Majorana fermion mode and also Majorana-Weyl fermionic mode 
for the optical cavity array  
with the relation between Rabi frequency oscillation and the
atom photon coupling strength. 
We have also presented the crossover between the Dirac fermion modes to
Majorana fermion modes and also the condition for appearence of Majorana-Weyl 
massless excitation mode. We have also presented several analytical relations
between the Majorana field, order and disorder operators.\\

Acknowledgement: The author would like to acknowledge the discussions
with Prof. S. Girvin  
during the international workshop/school on Dirac Materials and 
Chandrashekar lecture at ICTS and
the library and LAMP group of Raman Research Institute.
The author would like to thank Dr. P. K. Mukherjee for reading the
manuscript carefully. Finally the author would like to acknowledge
the DST (SERC) project and HRI Library.\\

{\bf Addendum}\\
{\bf Dirac fermion mode and analytical relation between fields:}\\

One dimensional quantum mechanical system both theoretically and
experimentally has studied extensively in last two decades with
several successful explanation for different physical systems.
There are several excellent research articles are
available in the literature. Here we discuss very briefly about
the Luttinger liquid physics of one-dimensional quantum many
body system.\\
The low energy excitation of a non-interacting system. In one
dimensional quantum mechanical system consists of Fermi points
at $\pm k_F $. The low energy fermionic state thus have momentum,
$ k \sim \pm k_F $ and a single particle energy close to Fermi energy
, $E_F $.
\beq
E (k) \simeq E_F + ( |k| - k_F ) v_F + ........ 
\eeq
Here we consider the one-dimensional infinite length system
and the elementary excitations across the Fermi points. Therefore
we can write the Fermi field,
\beq
{\psi}_s (x) = \int \frac{dk}{2 \pi} {\psi}_s (k) e^{i k x}
\eeq  
Here we consider the modes of momentum expansion in a neighbourhood 
of $\pm k_F $ of width $2 \Lambda $. Therefore, we can write the
fermionic field,
\beq
{\psi}_s (x) \simeq \int_{-\Lambda}^{\Lambda} \frac{dk}{2 \pi} 
e^{i (k+k_F)x } \psi_s (k + k_F) + \int_{-\Lambda}^{\Lambda} \frac{dk}{2 \pi} 
e^{i (k-k_F)x } \psi_s (k - k_F)
\eeq
One can write this analytical expression as,\\
\beq
{\psi}_s (x) \simeq e^{i k_F x} {\psi}_{s,R} (x) + e^{-i k_F x} {\psi}_{s,L},
\eeq
where ${\psi}_{s,R} (k) = {\psi}_s  (k+k_F) $ and
${\psi}_{s,L} (k) = {\psi}_s  (k-k_F) $.
The free fermion Hamiltonian,\\
\beq
H_0 = \sum_{s} \int \frac{dk}{2 \pi} \epsilon (k) {\psi_s}^{\dagger} (k) 
{\psi_s} (k) 
\eeq
\beq
H_0 = \sum_{s} \int_{-\Lambda}^{\Lambda} \frac{dk}{2 \pi} (k v_F )
( {{\psi}_{s,R}}^{\dagger} (k) {{\psi}_{s,R}} (k) -
{{\psi}_{s,L}}^{\dagger} (k) {{\psi}_{s,L}} (k) ). 
\eeq
Here we use $ {\epsilon}_k = k v_F $ near the Fermi points and
$ {\psi_{s} }^{\dagger} (x) = ( { {\psi}_{s,R}}^{\dagger} (x), 
{ {\psi}_{s,L}}^{\dagger} (x) ) $.
\beq
H_{0} = \sum \int \frac{dk}{2 \pi} {\psi_s}^{\dagger} (k) \sigma_3 k v_F  {\psi}_{s} (k). 
\eeq
\beq 
H_{0} = \sum_s \int dx {\psi_s}^{\dagger} (x) \sigma_3 k i v_F  {\partial_x} {\psi}_{s} (k), 
\eeq
where
$ 
 {\sigma}^{3} = \left (\begin{array}{cc}
      1 & 0 \\
    0  & -1
        \end{array} \right ) $.
We can define scalar field ($ \phi (x)$ ) and dual field ($\theta (x) $) by the
following relations\\
$ \phi (x) = \phi_R (x) + \phi_L (x) $, $\theta (x) = - {\phi}_R (x) + {\phi}_L (x) $. 
One canwrite the Dirac fermion density 
$\rho (x) = \rho_R (x) + \rho_L(x) = - \frac{1}{\sqrt{\pi}} {\partial_x} {\phi}_x $ and
the corresponding current density 
$ j(x) = v ( {\rho}_R - {\rho}_L = \frac{v_F}{\sqrt{\pi}} {\partial_x} {\theta} (x) $.
The commutattion relation between the scalar field (${\phi}$) and its dual field ($ \theta$).\\
$ [\phi (x1), \phi (x2)] = 0 = [\theta (x1), \theta (x2)] $, 
$ [\phi (x1), \theta (x2)] = i~sign (x2 - x1) $, $ [\phi (x1 ), \Pi (x2) ] = i \delta (x1 - x2) $.
Where $ \Pi (x) = \partial_x \theta (x) $. The corresponding Lagrangian debnsity for massless
Dirac fermions is\\
\beq
L = i {\psi_R}^{\dagger} (\partial_t + v_F \partial_x ) {\psi_R} +
 i {\psi_L}^{\dagger} (\partial_t - v_F \partial_x ) {\psi_L} 
\eeq 
Now we present the algebra of the operators, $ e^{\pm \phi (x)} $ and 
$ e^{\pm \theta (x)} $. \\
\begin{equation}
e^{i \alpha \sqrt{\pi} \phi (x1 )} e^{i \beta \sqrt{\pi} \theta (x2 )} =  
e^{i \beta \sqrt{\pi} \theta (x2 )} e^{i \alpha \sqrt{\pi} \phi (x1 )} ~sign (x2 -x1)   
\end{equation}
\begin{equation}
e^{i \alpha \sqrt{\pi} \phi (x1) } {\psi_{R/L}} (x2) = {\psi_{R/L}} (x2) 
e^{i \alpha \alpha \sqrt{\pi} \phi (x1) }~sign(x2 - x1) 
\end{equation} 
\begin{equation}
e^{i \beta \sqrt{\pi} \theta (x1) } {\psi_{R/L}} (x2) = - {\psi_{R/L}} (x2) 
e^{i \beta \sqrt{\pi} \theta (x1) }~sign(x2 - x1) 
\end{equation} 

{\bf Analytical relations between order and disorder operators:}\\ 
The analytical relation between the Pauli operators and $\mu $
operators are relagated to the appendix B of this manuscript, please put the
following analytical expression in the appendix:\\
\beq
 {{\mu}_z }^2 = 1 = {{\mu}_x }^2 ,
\eeq
\beq
{{\mu}_z } (n -1/2 ) {{\mu}_z } (n +1/2 ) = {\sigma}_x (n).
\eeq
\beq
{{\mu}_x } (n + 1/2) = {\sigma}_z (n) {\sigma}_z (n+1),
\eeq
\beq
{{\mu}_z } (n + 1/2) = \Pi_{j=1}^{n}  {\sigma_x} (j).
\eeq
\beq 
{{\sigma_z}} (n) = \Pi_{j=0}^{n-1} {\mu_x} (j+1/2) ,
\eeq
\beq
 [{\mu}_x (n + 1/2) , {\mu}_z (n^{'} + 1/2) ]= 2 \delta_{n,n^{'} }
\eeq
\beq 
 [{\mu}_z (n + 1/2) , {\mu}_z (n^{'} + 1/2) ]= 0 ,
\eeq
\beq
 [{\mu}_z (n + 1/2) , {\sigma}_x (n^{'} ) ]= 0
\eeq 
The operator ${\mu}_z  (n+1/2 )$ acting on the original spin of the lattice
makes a spin flip of all those spin placed on the left hand side of spin 
at the site n. Therefore ${\mu}_z (n +1/2 ) $ is a kink operator, it introduce
the disorder in the system.
It is very clear from the above analytical relation of the operators that
Here we express the analytical relation between the Majorana
operators and the disorder and Pauli operators:\\
\beq
{\chi}_1 (n) = {\sigma}_z (n) {\mu}_z (n+1/2) = - {\mu}_z (n+ 1/2) {\sigma}_z (n)  
\eeq
\beq
{\chi}_2 (n) = {\sigma}_z (n) {\mu}_z (n-1/2) =  {\mu}_z (n- 1/2) {\sigma}_z (n)  
\eeq
\beq
{\sigma}_z (n) {\chi}_2 (n) =  {\mu}_z (n-1/2) =  {\chi}_2 (n) {\sigma}_z (n)  
\eeq
\beq
{\sigma}_z (n) {\chi}_1 (n) =  {\mu}_z (n+1/2) = - {\chi}_1 (n) {\sigma}_z (n)  
\eeq
\beq
{\sigma}_z (n)=  {\mu}_z (n-1/2) {\chi}_2 (n) =  {\chi}_2 (n) {\mu}_z (n- 1/2)   
\eeq
\beq
{\mu}_z (n+1/2) {\chi}_1 (n) = - {\chi}_1 (n) {\mu}_z (n+1/2)= {\sigma}_z (n)  
\eeq

{\bf Derivation of Dirac equation for Majorana fermion field} \\
The equation of motion for the ${\sigma}_z (n)$ is the following:\\
\beq
\frac{\partial \sigma_z (n)}{\partial \tau} = [H, \sigma_z (n)]= {\sigma_x (n) \sigma_z (n) }
\eeq 
The equation of motion for $\mu_z (n+1/2) $ is the following:\\
\bea
\frac{\partial \mu_z (n+1/2)}{\partial \tau} & = & 
\lambda {\mu_x (n +1/2 ) \mu_z (n + 1/2) } \non\\
& & = \lambda \sigma_z (n) \sigma_z (n + 1/2) \mu_z (n + 1/2) \\  
\eea
Now we use the properties
of the $\sigma$ and $\mu$ operators to derive the equation of
motion for the Majorana fields $ \chi_1 (n) $ and $\chi_2 (n) $.\\
\beq
\frac{ \partial \chi_1 (n) }{d \tau} = \frac{\partial \sigma_z (n)}{\partial \tau}
\mu_z (n+ 1/2) + \sigma_z (n) \frac{\partial \mu_z (n)}{\partial \tau}.
\eeq 
\beq
\frac{ \partial \chi_1 (n) }{d \tau} = \sigma_x (n) \sigma_z (n) 
\mu_z (n+ 1/2) + \lambda \sigma_z (n) \sigma_z (n) \sigma_z (n+1) \mu_z (n + 1/2). 
\eeq
\bea
\frac{ \partial \chi_1 (n) }{d \tau} & = & - \sigma_z (n) 
\mu_z (n- 1/2) \mu_z (n+ 1/2) \mu_z (n+ 1/2) \non\\ 
& & + \lambda \sigma_z (n) \sigma_z (n) \sigma_z (n+1) \mu_z (n + 1/2). 
\eea
\beq
\frac{ \partial \chi_1 (n) }{d \tau} = - \chi_2 (n) 
+ \lambda \chi_2 (n + 1). 
\eeq
Now the equations of motion for $\chi_2 (n) $ are
\bea
\frac{ \partial \chi_2 (n) }{d \tau} & = & \frac{\partial \sigma_z (n) }{\partial \tau}
\mu_z (n -1/2) \non\\ 
& & + \sigma_z (n) \frac{\partial \mu_z (n - 1/2) }{\partial \tau} 
\eea
\bea
\frac{ \partial \chi_2 (n) }{d \tau} & = & \sigma_x (n) \sigma_z (n) 
\mu_z (n -1/2) \non\\ 
& & + \lambda \sigma_z (n) \sigma_z (n-1) \sigma_z (n) \mu_z (n - 1/2) 
\eea
\bea
\frac{ \partial \chi_2 (n) }{d \tau} & = & \mu_z (n-1/2) \mu_z (n + 1/2) \sigma_z (n) 
\mu_z (n -1/2) \non\\ 
& & + \lambda \sigma_z (n-1) \mu_z (n - 1/2). 
\eea
After a little bit of calculations and using the relation between the
disorder operators (Eq. 23-30), we finally arrive the equation of motion of $\chi_2 (n) $
as,
\beq
\frac{ \partial \chi_2 (n) }{d \tau} = -\chi_1 (n) 
 + \lambda \chi_1 (n-1). 
\eeq
These two fields, $\chi_1 (n)$ and $\chi_2 (n) $ satisfy the following relations,
$ \{ \chi_1 (n1) , \chi_2 (n2) \} = 2 \delta_{n1, n2} $. One can write down
the above equation in the following compact form, \\
\beq
( {\gamma}^0  \frac{\partial }{\partial t} + {\gamma}^{3} \frac{\partial}{\partial r}
+ m ) \chi =0 . 
\eeq
where ${\chi}^{\dagger} = ( \chi_1 , \chi_2 )$
and $ m= \frac{1 - \lambda}{\alpha} $, 
$ {\gamma}^{0} = \left (\begin{array}{cc}
      0 & 1 \\
    1  & 0
        \end{array} \right ) ,
 {\gamma}^{3} = \left (\begin{array}{cc}
      1 & 0 \\
    0  & -1
        \end{array} \right ) $
.\\
\end{document}